\documentclass[10pt]{forma}

\usepackage{graphicx}
\usepackage{wrapfig}

\begin{document}
\hspace{1.0cm} \parbox{15.0cm}{

\baselineskip = 15pt

\noindent {\bf Abnormal lithium abundance in several Ap-Bp stars}

\bigskip
\bigskip

\noindent {\bf A.~Shavrina$^{1}$, N.S.~Polosukhina$^{2}$,
N.A.~Drake$^{3}$, D.O.~Kudryavtsev$^{4}$, V.F. Gopka$^{5}$, V.A. Yushchenko$^{5}$,
A.V.~Yushchenko$^{6}$}

\bigskip

\baselineskip = 9.5pt
\noindent {\small \copyright~2009}
\smallskip

\noindent {\small {\it $^{1}$Main Astronomical Observatory NAN Ukraine,
$^{2}$Crimean Astrophysical Observatory, Ukraine,
$^{3}$Sankt-Peterburg University, Russia,
$^{4}$Special Astrophysical Observatory, Russia,
$^{5}$Odessa University, Ukraine,
$^{6}$Sejong University, (South) Korea
}} \\
\noindent {\small {\it e-mail:}} {\tt shavrina@mao.kiev.ua$^{1}$,
polo@crao.crimea.ua$^{2}$,
drake@on.br$^{3}$,
dkudr@sao.ru$^{4}$,
gopkavera@mail.ru$^{5}$, 12345678e@mail.ru$^{5}$,
yua@sejong.ac.kr$^{6}$}

\baselineskip = 9.5pt \medskip

\medskip \hrule \medskip

\noindent

Using observations of rapidly rotating roAp star HD 12098, obtained with 6m
 telescope of SAO RAS  in 2005-2008 in nine rotational phases (0.001-0.817),
and code V. Tsymbal ROTATE we identified 3 lithium spots on the surface of star,
one of them coincided with the position of one magnetic pole and two
 others are located near the other pole, symmetrically with respect to it.
 Perhaps magnetic field of the star has a more complex structure than the dipole
 \cite{Shavrina2008a}. Lithium abundance in the spots is greater than cosmic
one. It  can be explained by diffusion of lithium in the presence of a magnetic
 field or by production of lithium in the reactions of "spallation" on the
surface of star near the magnetic poles (see, for example, \cite{Goriely2007}).
Enhanced lithium abundances was also determined for other two slowly rotating
Ap stars HR~465 and HD~965. The analysis is performed using a synthetic spectrum
method taking into account magnetic splitting of lines by the SYNTHM code of
S. Khan  \cite{khan04}.
\medskip \hrule \medskip
}

\vspace{1.0cm}

\baselineskip = 11.2pt

\noindent {\small {\bf INTRODUCTION}}
\medskip
\noindent

Spectral observations of Ap-CP stars at BTA telescope (SAO RAS) with echelle
spectrometer NES \cite{Pan02} are performed under the project "Lithium in CP
stars".  It was found several stars with anomalous lithium content. The
oscillating star HD~12098, first roAp star, opened recently on northern sky
\cite{girish}, was among these stars.
 The spectrum of this star showed strong and variable Li I line
 6708 \AA, analysis of which revealed a high abundance of lithium
 \cite{Shavrina2008}. It has been shown the presence of lithium spots on the
surface of the star \cite{Shavrina2008a}, similar to two roAp stars HD~83368
and HD~60435, for which lithium spots were previously detected on their surface
 \cite{Shavrina2001}.
 In 2008 three new phases have added to the observations  and we made a new
analysis based on 9 observed phases (0.001-0.817) with Tsymbal's code ROTATE.
 Also the lithium blend 6708 \AA~   was analyzed in the spectra of two
slowly rotating Ap stars HR~465 and HD~965, for which we used the method of
synthetic spectrum taking into account the magnetic splitting of lines
\cite{khan04}.

\medskip

\noindent {\small {\bf The star HD~12098}}
\medskip

\noindent

The method of synthetic spectrum on the base of model atmospheres was applied
for analysis taking into account the VALD list and some additional lines of
rare earth elements (REE), calculated by the authors \cite{Shavrina2003}.
Basing on Fe I/Fe II lines we selected  the model atmosphere for HD~12098
from Kurucz' grid: Teff = 7750~K and lg g = 4.0.
Using this effective temperature, evolutionary tracks from the work of Schaller
et al. \cite{schaller92} and the formula from the paper of Kochukhov and Bagnulo
\cite {koch06}, we determined the stellar radius R = 1.71 $ R_\odot$, log g =
4.2, Vrot = 15.8 km/s, i = 55$^o$ and vsini = 13 km/s.
The profile of the lithium blend Li I 6708 \AA\ were calculated using
the code SYNTHM of S.Khan \cite{khan04} taking into account magnetic splitting.
In the work of \cite{Shavrina2008a} a model of the dipole magnetic field for
HD~12098 was constructed on the base of Ryabchikova et al. \cite{Ryabchik05}
values  of effective magnetic field for this star. With the code ROTATE of
V.Tsymbal (see \cite{Shavrina2001}) the observations of this star in 2005-2007
(phases 0.018 - 0.573, 6 phases) were analyzed. Lithium spots, that almost
coincided with the locations of the poles of magnetic dipole, have been
identified and the lithium abundance in them were determined. Observations
of 2008 (DK) added 3 new phases and we have undertaken a new analysis by the
code ROTATE basing on 9 observed phases (0.001-0.817). Instead of two lithium
spots \cite{Shavrina2008a},  new analysis showed the presence of 3 lithium
spots, one of which coincides with the position of one pole of the magnetic
field, the two other spots are located near the other pole, symmetrically with
respect to it.
Possibly magnetic field of the star has a more complex structure than the dipole
\cite{Shavrina2008a} and the magnetic field model must be recalculated with new
observations.
Table 1 shows longitude and latitude of the Li spots in equatorial coordinates,
radius of the each spot and the lithium abundance in the spot $\epsilon $(Li).
The observed profiles Li I 6708 \AA\ line for the 9 rotational phases and model
profiles (calculated with spot parameters of Table 1) are shown in Figure 1.
Lithium abundance in the spots is higher than cosmic. It can be explained by
diffusion of lithium in the presence of a magnetic field or by production
of lithium in the reactions of "spallation" on the stellar surface near the
poles of the magnetic field (see, for example, \cite{Goriely2007}).

\begin{table}
\begin{center}
\caption{\label{abcd} Lithium spots parameters}
\begin{tabular}{|c|c|c|c|c|}
\hline
\noalign{\smallskip}

N & Longitude & Latitude & R &  $\epsilon$(Li) \\

\noalign{\smallskip}
\hline
\noalign{\smallskip}
1 & 30$^o$ & -20$^o$ & 40$^o$ & 5.0 \\
2 &180$^o$ &  25$^o$ & 70$^o$ & 4.2 \\
3 &290$^o$ & -25$^o$ & 40$^o$ & 4.4 \\
\noalign{\smallskip}
\noalign{\smallskip}
\noalign{\smallskip}
\hline
\noalign{\smallskip}
\end{tabular}
\end{center}
\end{table}

\begin{figure}
\begin{center}
\includegraphics[width=80mm, height=80mm, angle=0]{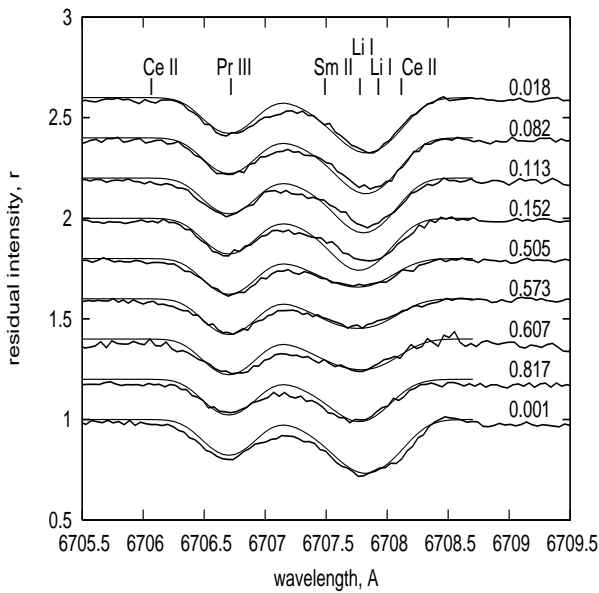}
\caption{\small  Observed (bold line) and model (7750/4.2) spectra
of HD 12098 in the range 6705.5 -- 6708.5 \AA\ for 9 rotational phases
(on the right)}
\label{figure:fig1}

\end{center}
\end{figure}

\medskip
\noindent {\small {\bf The star HD~465}}
\medskip

\noindent

The star HR~465 (HD~9996) shows the variability of the spectrum with 23 years
period (\cite{Preston70}). The spectra observations were made by AYu
in 2004 with eshelle spectrograph BOES $(http://www.boao.re.kr/BOES/)$ at 1.8m
telescope at the Bogunsan observatory (BOAO) in Korea (R = 80000, $S/N>$=150,
the spectral region 3780--9500 \AA). In addition, we used the spectrum of this
star from spectral databases ELODIE - http://www.obs-hp.fr/archive/archive.html).
These spectra are separated by approximately of half-period of rotation, in the
so-called phase of chrome, with enhanced lines of chromium (ELODIE) and the
phase of rare earth elements (REE) with enhanced lines of REE (BOAO). Using
profiles of ionized iron at 6148 and 6150 \AA\ we determined the value of the
surface magnetic field $ B_s $ in two phases, which were needed to model
the profile Li I 6708 \AA.~~ These values  were 1500-2000 G for BOES spectrum
and 4500-5000 G for ELODIE one (Figs. 2, 3)
The effective temperature and the gravity for the star HR~465 were
identified in  the work of V.Yushchenko et al. \cite{yu08} on the base of
equivalent widths of neutral and ionized iron lines:
Teff = 11840 K, lg g = 4.3. To simulate the BOES spectrum, which shows
strong  REE and lithium lines, we used model atmosphere from Kurucz's grid with
close parameters 11750/4.0 \cite{Kur94}, a list of lines VALD \cite {vald} and
some additional REE lines, calculated by the authors \cite{Shavrina2003}. We
determined the lithium abundance $\epsilon$ (Li) = 5.40, which is 2 orders
exceeds the "cosmic" lithium abundance, the abundance of cerium $\epsilon$(Ce)
= 6.99 and upper limit of samarium abundance $\epsilon $(Sm) = 6.55. The last
two values are significantly (more than 5 orders of magnitude) higher than the
solar abundance of cerium and samarium. Note that to describe the profile of
the Fe II 6708.885 \ AA \ we have forced to use a resolution in the spectrum of
R = 50000 instead of 80000 and the iron abundance $\epsilon$(Fe) = 8.90, which
is more than one order of magnitude above solar (Fig. 4). In the Figure 4 we see
the doubling of lithium and REE lines, apparently due to the spots structure
of the stellar surface.

\begin{figure*}[t]
\includegraphics[angle=0,scale=0.45]{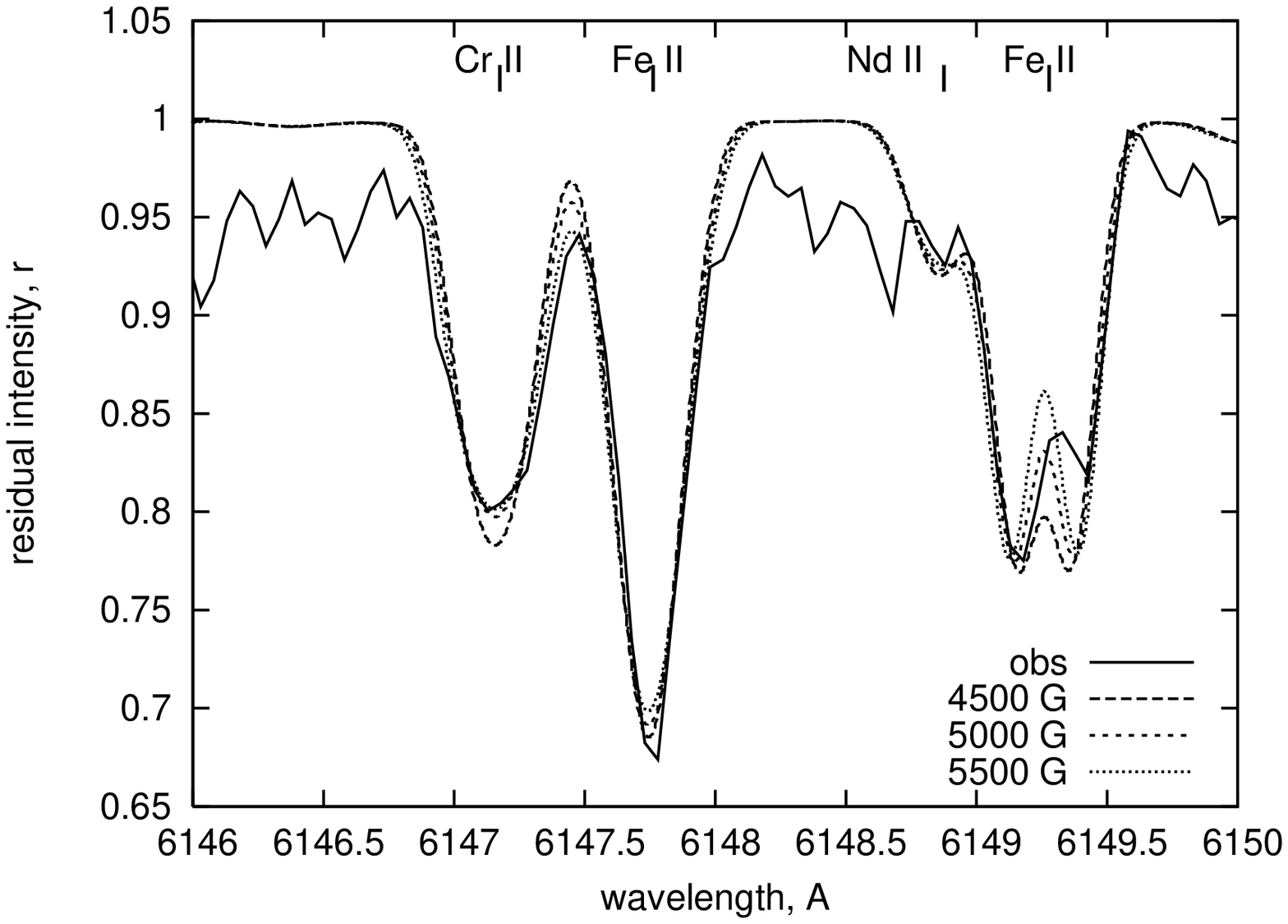}
\includegraphics[angle=0,scale=0.45]{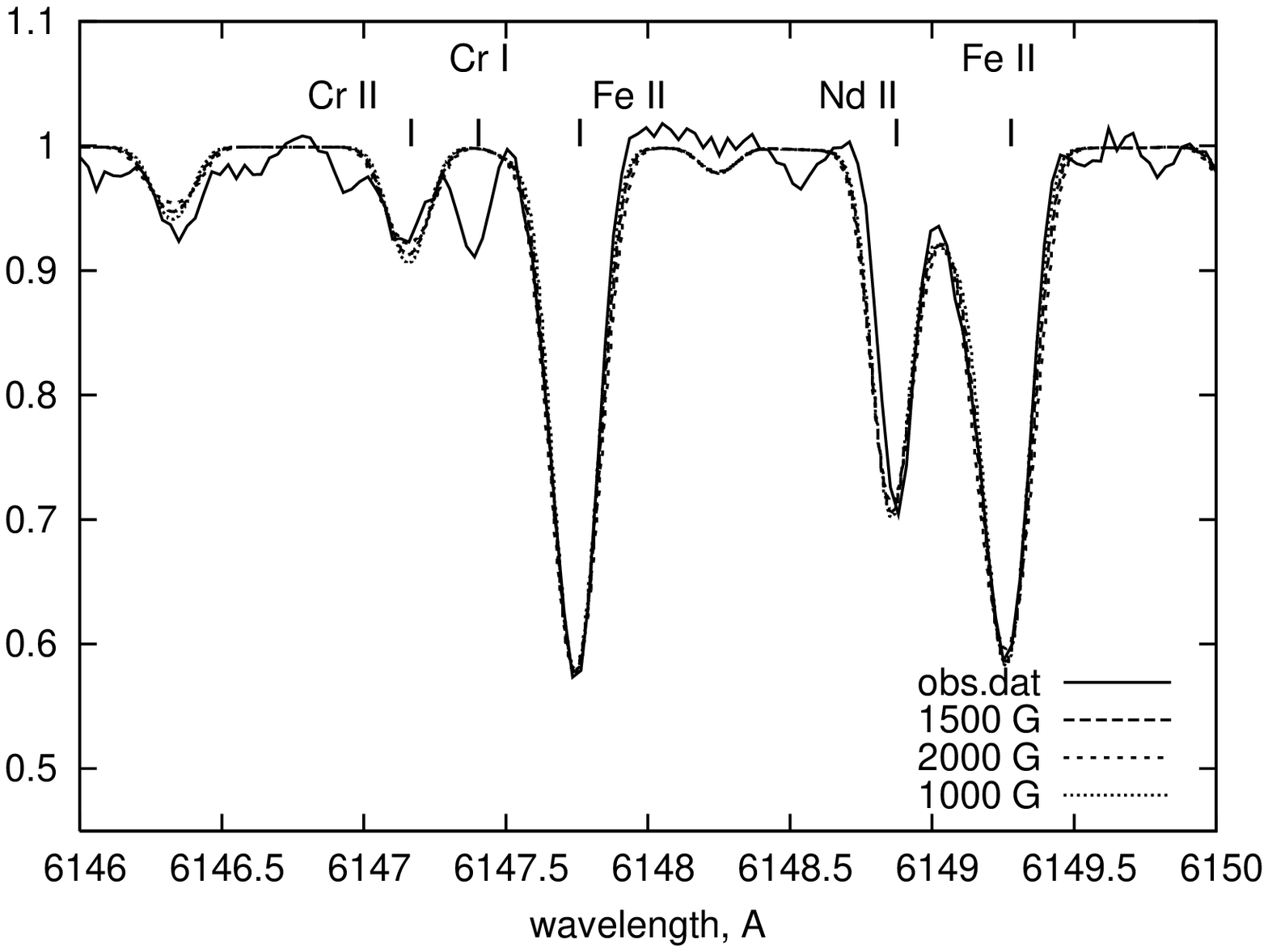}
\caption{ a) Observed spectrum (bold line) in the phase of "chrome" and
model(11750/4.0) spectra of HR 465 in the range 6147-6150 \AA\
b) Observed spectrum (bold line) in the phase of "REE" and model(11750/4.0)
spectra of HR 465 in the range 6147-6150 \AA\
}
\label{mag-fe}
\end{figure*}

\medskip
\noindent {\small {\bf The star HD~965}}
\medskip

\noindent

The observed spectrum of magnetic peculiar star HD 965 of spectral class A5p
(SIMBAD, http://simbad.u-strasbg.fr/simbad/) was obtained by AYu with eshelle
spectrograph BOES at 1.8m telescope at the Bogunsan observatory (BOAO) of Korea
in December 2004. HD 965 - is slowly rotating Ap star with strong magnetic
fields: surface magnetic field Bs = 4400 G has been measured by Mathys et al.
1997 (\cite{Mathys97}. The rotational period of the star is more than 10 years
\cite{Elkin05}.
Parameter set Teff = 7500, lg g = 4.0 and vsini = 3 km/s  needed to calculate
model profile of Li I 6708 \AA\ blend were taken from \cite{Ryabchik08}.
The method of synthetic spectrum on the base of the model atmosphere 7500/4.0
from Kurucz grid \cite{Kur94} taking into account the lines of VALD list and
some additional REE lines, calculated by the authors \cite{Shavrina2003},
was applied to simulate profiles of blend 6708 \AA\, composed of the lithium
doublet 6707.8  and Ce II 6708.1 \ AA \ lines. The magnetic splitting
of lines was handled with Khan's code SYNTHM \cite{khan04}.
Comparing the observed spectrum of the star HD 965 with computed ones
we determined the boundaries of the abundance of lithium and cerium,
namely $\epsilon $(Li) = 2.90 - 3.09, which are close to the "cosmic" lithium
abundance. Apparently, the strong magnetic field prevent the mixing and the
destruction of the original lithium.
The cerium abundance is in the range $\epsilon$(Ce) = 3.85 - 4.27, which is
almost three order greater than solar. Note that the lines of both elements in
spectrum of this star show spots structure apparently associated with
two poles of the magnetic field.

\begin{figure*}[t]
\includegraphics[angle=0,scale=0.45]{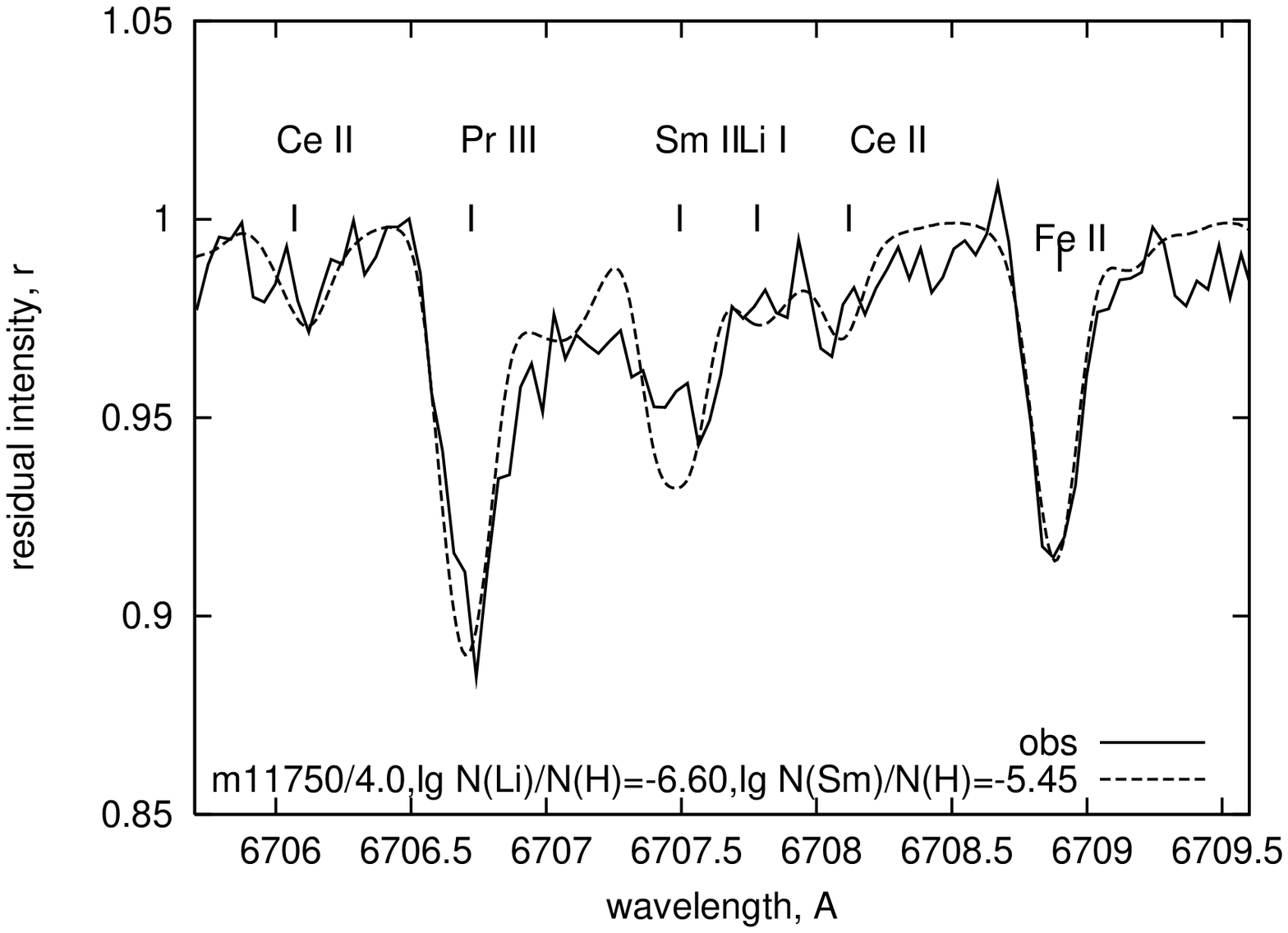}
\includegraphics[angle=0,scale=0.45]{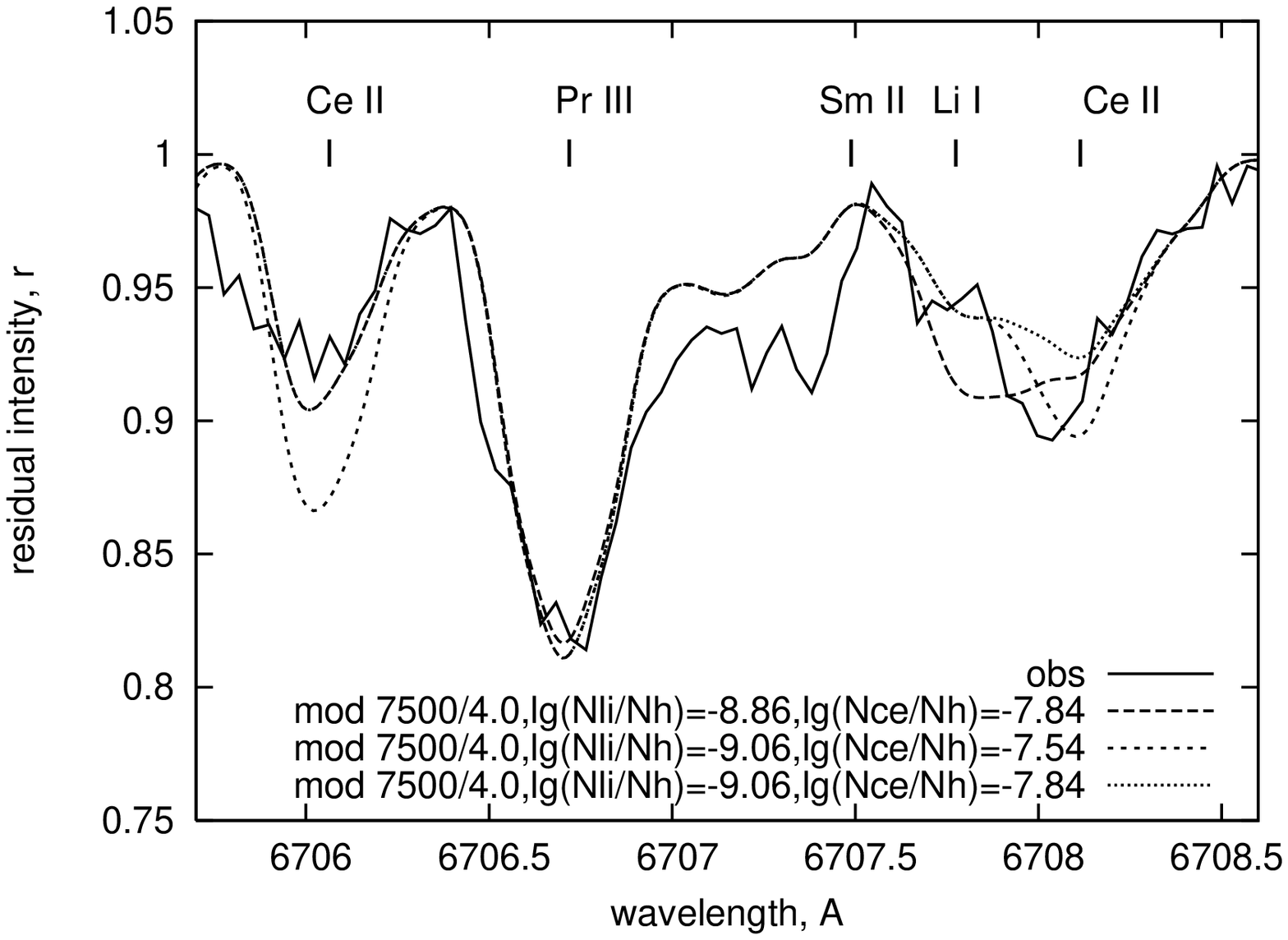}
\caption{ a) Observed spectrum (bold line) in the phase "REE" and model spectra
(11750/4.0) of HR 465 in the range 6705.5 - 6708.5 \AA.~~
b) Observed (bold line) and model (7500/4.0) spectra of  HD 965 for 6705.5 -
6708.5 \AA~ for boundary abundances of Li and Ce (see figure).
}
\label{mag-fe}
\end{figure*}

\medskip
\noindent {\small {\bf CONCLUSION}}
\medskip

\noindent
A new analysis of the spectra of roAp star HD~12098 recently discovered in the
northern sky shows the presence of 3 lithium spots, one of which coincides with
the location of one magnetic pole, the two other spots are near another pole,
symmetrical for him. Perhaps magnetic field of the star is more complicated,
than the dipole (\cite{Shavrina2008a}) and its model must be recalculated,
taking into account new observations.
Enhanced lithium abundances in the spots on the surface of this rapidly rotating
star and in the atmospheres of the two other slowly rotating Ap stars -
HR~465 and HD~965 can be caused by physical processes that prevent
mixing of stellar matter in the atmospheres and preserve their original
high levels, namely, the suppression of convection by magnetic field and
coupled with diffusion of lithium, directed to the surface of stars.
Another possibility is the real processes of lithium production in the surface
layers of the atmosphere, such as the reactions of "spallation" in which
accelerated in the magnetic fields protons and alpha-particles destroy the
heavier atoms C, N, O with formation of lithium. This possibility is shown in
the theoretical work of Goriely \cite{Goriely2007} for Przybylski's star
(roAp star with a large excess of REE).
\\

{\bf Acknowledgements}
 We thank V. Tsymbal and S. Khan for the opportunity to use their
 codes ROTATE and SYNTHM.

\medskip
\noindent {\small {\bf REFERENCES}}
%

\end{document}